# Band-like electron transport in organic transistors and implication of the molecular structure for performance optimization

By *Nikolas A. Minder*, *Shimpei Ono*, *Zhihua Chen*, *Antonio Facchetti\**, and *Alberto F. Morpurgo\**

[*]     N. A. Minder, Dr. S. Ono, Prof. A. F. Morpurgo
DPMC and GAP, University of Geneva
24 quai Ernest-Ansermet,
CH - 1211 Geneva (Switzerland)
E-mail: alberto.morpurgo@unige.ch
        Dr. S. Ono
Central Research Institute of Electric Power Industry
Komae, Tokyo 201-8511 (Japan)
        Dr. Z. Chen, Prof. A. Facchetti
Polyera Corporation, 8045 Lamon Avenue
Skokie, IL 60077 (USA)
E-mail: afacchetti@polyera.com
        Prof. A. Facchetti
Department of Chemistry and Material Research Center
Northwestern University, 2145 Sheridan Road,
Evanston, IL 60208 (USA)

Understanding the microscopic processes limiting the charge carrier mobility in organic field-effect transistors (OFETs)[1] is as important as to improve the quality of existing devices.[2-10] To this end, OFETs based on single-crystals[11] have considerable potential, owing to their unprecedented structural quality and chemical purity. Single-crystal transistors have been used to demonstrate the occurrence of band-like transport in OFETs, through the observation of an (anisotropic[12]) carrier mobility increasing with decreasing the temperature[13-15] and of the Hall effect.[16,17] They have also led to a detailed microscopic understanding of how transport is influenced by the gate dielectric,[18-20] which has a dominant effect in determining the field-effect mobility that is measured experimentally.[18,21] Despite these successes, our understanding of OFET charge transport remains limited:[22] only a few materials exhibit band-like transport in an OFET configuration, in all cases corresponding to *p*-channel devices. More importantly, it is not understood why only those materials exhibit band-like transport. At this stage, it is crucial to broaden the class of organic semiconductors in which band-like transport is observed, and to identify mechanisms and properties – common to all of these materials – that favor its occurrence.

Recently, we have shown that single-crystal OFETs based on the molecule *N,N'*-bis(n-alkyl)-(1,7 and 1,6)-dicyanoperylene-3,4:9,10-bis(dicarboximide)s (PDIF-CN$_2$[23], see **Fig. 1a**) – where the presence of fluoroalkyl and cyano substituents increases the core electronic affinity[24,25] – exhibit significantly improved characteristics with respect to the majority of *n*-channel OFETs studied previously.[26,27] These characteristics include very high electron mobility values, great reproducibility, and stable ambient operation, which make PDIF-CN$_2$ a candidate to carry out fundamental studies of electron transport. Here we investigate PDIF-CN$_2$ single-crystal FETs, demonstrate band-like electron transport in a transistor configuration, and investigate the influence of the gate dielectric on the transport properties. The analysis of our observation, in conjunction with recent studies of band-like transport in different organic *p*-channel devices, indicates that charge transport in organic FETs is not only affected by the gate insulator dielectric properties but also by dielectric environment due to the polarizability of the π-conjugated core intrinsic to the organic semiconductor.

PDIF-CN$_2$ transistors were fabricated by lamination of thin single crystals, grown by vapor transport, onto different types of substrates. Elastomeric poly(dimethylsiloxane) (PDMS) substrates with a recessed gate[12,27,28] were used to realize devices in which the FET channel is not directly in contact with a dielectric material (see Fig. 1b – we refer to these devices as to "vacuum-gap transistors"). Additionally, Si/SiO$_2$ substrates coated with either Cytop ($\varepsilon$=2.1) or PMMA ($\varepsilon$=3.8) were also used to explore how the dielectric constant of the gate insulator affects electron conduction. Cytop and PMMA were chosen because, in contrast to non-polymeric dielectrics such as inorganic oxides, no chemical groups acting as electron traps are present on their surface.[29]

Figs. 1c and 1d show the transfer and output characteristics for a 25 μm vacuum-gap single-crystal PDIF-CN$_2$ OFET. The linearity of the $I_{DS}$-$V_{SD}$ plots at low bias and the absence of any noticeable *I-V* hysteresis indicate the high quality of these transistors. The FET mobility $\mu = (L'/W)(1/C_i \cdot V_{XX})(dI_{DS}/dV_{GS})$ was extracted from trans-conductance measurements in a four-terminal configuration – $V_{XX}$ is the voltage between the two probes separated by a distance *L'*, W the crystal width, $C_i$ the capacitance per unit area of the gate dielectric, and $V_{GS}$ the gate voltage. For a total of over 20 vacuum-gap devices, the room-temperature *μ* values are between 2.1 - 5.1 cm$^2$/Vs (average mobility $\bar{\mu}_{Vac} = 3.68 \pm 0.98$ cm$^2$/Vs). Representative transfer and output characteristics for FETs based on Cytop and PMMA gate dielectrics, demonstrating the high quality of the devices, are shown in **Fig. S1** of the supporting information. The room-temperature mobility values are in the range 2.2 - 3.1 cm$^2$/Vs ($\bar{\mu}_{Cytop} = 2.55$ cm$^2$/Vs $\pm 0.42$ cm$^2$/Vs) and 1.7 - 2.3 cm$^2$/Vs ($\bar{\mu}_{PMMA} = 2.07 \pm 0.23$ cm$^2$/Vs) in the two cases, respectively. These data indicate that the electron mobility decreases with increasing the relative dielectric constant (ε) of the gate insulator from 1 (vacuum) to 3.8 (PMMA), similarly to what is observed in rubrene devices. The magnitude of the effect, however, is smaller in PDIF-CN$_2$ than in rubrene, where the mobility decreases from ~ 20 cm$^2$/Vs to ~ 5 cm$^2$/Vs when going from vacuum-gap to a gate dielectric with ε ~ 4 (such as PMMA).[18]

We studied the temperature dependence of the electron mobility for the three gate dielectrics and observed an increase in electron mobility while cooling from room temperature to a characteristic temperature, *T\**, whose value depends on the gate insulator, and then decreases upon further cooling. **Fig. 2** shows $\mu_{FET}$-*T* data from representative devices. *T\** is around 200 K for vacuum-gap devices (red circles), 225 K for FETs on Cytop (blue triangles) and 260 K on PMMA (green squares). The behavior is analogous to that of holes in rubrene single crystal FETs, in which case $T^*$ is found to range typically between 150 and 200 K when vacuum is used as gate dielectric.[16,19] On the best vacuum-gap devices we have also measured the Hall effect[16,17] at room $T$ (**Fig. 3c**). The density of electrons ($n_H = I_{DS}B/eV_{XY} \approx$ 3.46·10$^{10}$ cm$^{-2}$) extracted from the slope of the Hall voltage as a function of the magnetic field $B$ (**Fig. 3d**) are in good agreement with the value estimated from the gate capacitance ($n_{FET} = C_i \cdot (V_G - V_{Th})/e \approx 3.31 \cdot 10^{10}$ cm$^{-2}$), indicating the consistency of our analysis. The unambiguous observation of the Hall effect together with a negative $d\mu/dT$ over a finite temperature range is a characteristic signature of charge carriers delocalized over a few molecules moving in a two-dimensional charge transport plane.[16,17] This is what is usually referred to as *band-like* transport, to distinguish it from true band transport where charge carriers are delocalized on a length scale much larger than the lattice constant. For *n*-channel devices, PDIF-CN$_2$ single crystal FETs are the first in which transport occurs in this band-like regime, indicating that their quality is comparable to that of the very best *p*-channel organic transistors.

Important information can be obtained by analyzing the dependence of transport in PDIF-CN$_2$ FETs on the dielectric constant of the gate insulator. For holes in *p*-channel OFETs, different mechanisms account for the dependence of the mobility on the dielectric properties of the gate insulator.[19-21,30,31] In the regime of strong coupling (large polarizability of the gate dielectric, as it occurs in high-ε oxides) it has been shown that this dependence can be understood in terms of the formation of interfacial polarons.[19,20] With polymeric dielectrics, as those used in our PDIF-CN$_2$ devices, the coupling is not strong enough[19] for these interfacial polarons to form. The dominant mechanism in the present case is the so-called dipolar disorder,[21,32] originating from the random orientation of the dipolar components of the polymeric dielectric. These dipoles generate potential fluctuations in the transistor channel which broaden the density of states of the organic semiconductors. For OFETs based on disordered materials, the broadening results in a lower density of states at the Fermi energy and causes a decrease of the hopping probability (and hence of the mobility).[21,33,34] Even though in PDIF-CN$_2$ single crystal FETs transport occurs in the band-like regime and is not mediated by hopping, dipolar disorder still affects the density of states in the organic semiconductor in the transistor channel, resulting also in this case in the decrease of the FET mobility.

In the presence of dipolar disorder, a random potential $\phi$ is generated by the dipoles in the polymer gate dielectric, which locally shifts the energy of the states in the band tail (i.e., shallow traps) in the FET channel. The potential $\phi$ is a Gaussian distributed random variable, $P(\phi) \propto \exp(-\phi^2 e^2 / 2\Delta^2)$, where $\Delta$ quantifies the magnitude of the dipolar disorder ($e$ is the elementary charge). The average density of states in the band tail is then given by $\tilde{\rho}(E) = \int_{-\infty}^{\infty} d\phi P(\phi) \rho(E - e \cdot \phi) \propto \tilde{\rho}_0 \cdot \exp\left(-\frac{E^2}{2(E_0^2 + \Delta^2)}\right)$, which, as compared to the distribution $\rho(E)$ of tail states in the absence of the dielectric, is broadened by an amount $\Delta$. The broadening corresponds to an increased depth of the energy distribution of the band tail states that causes a lowering of the FET mobility and a narrowing of the temperature range in which the mobility is seen to increase upon lowering temperature (see supporting information for details). This is precisely what is observed in our experiments, when comparing devices with vacuum, Cytop, and PMMA as gate insulators (Fig. 2d).

To estimate the role of the dielectric dipolar disorder in the band-like transport regime we use a simple phenomenological model which we have developed to analyze the behavior of TMTSF single-crystal FETs[14] (for details see supporting information). With this model, we fit the measured mobility temperature dependence for the FETs based on the three dielectrics and extract the width of the band tail (Fig. 2d). The result obtained for the vacuum-gap dielectric, where the dielectric dipolar disorder is absent, gives the energy width $E_0$ of the band tail in PDIF-CN$_2$ crystals (38 meV) due to the disorder present in the semiconductor only. For the devices with Cytop and PMMA gate dielectrics, $E_{Cytop} \approx 43$ meV and $E_{PMMA} \approx 56$ meV, respectively, include the effect of dipolar disorder, whose magnitude is given by $\Delta = \sqrt{E_{Cytop/PMMA}^2 - E_0^2}$. We find $\Delta \approx 20$ meV for Cytop and $\Delta \approx 41$ meV for PMMA, comparable to – but smaller than – the values estimated by Veres *et al.* (37 meV for Cytop and 70 meV for PMMA[21]) for the same gate dielectrics. The smaller magnitude of the potential fluctuations obtained in the case of PDIF-CN$_2$ FETs is expected due to the presence of the fluorocarbon substituents, which spatially separate the device channel from the gate dielectric. Indeed, it can be easily estimated that the core substituents roughly double the distance between the dipoles at the surface of the dielectric and the charge carriers, as compared to the case where no substituents are present: given the slow spatial dependence of Coulomb potential and the molecular size of the dipoles responsible for the disorder, the observed suppression of $\Delta$ has the correct order of magnitude.

Finding that the molecular structure has a large effect on suppressing coupling between the charge carriers in the channel and the electrical polarization in the gate dielectric provides an important indication as to the microscopic mechanisms that affect the carrier mobility in organic FETs. PDIF-CN$_2$ single crystals consist of alternating π-conjugated perylene and insulating fluorocarbon layers, an arrangement that is typical for several high-mobility organic semiconductors. The fluorocarbon chains linked to the perylene core do not only screen the electrons accumulated in the first π-conjugated layer (i.e., the FET channel) from the gate insulator, but also from the nearby π-conjugated layers in the crystal, that are equally effective in determining the dielectric environment through their polarizability. Indeed in devices with vacuum as gate insulator the dielectric environment seen by the charge carriers is entirely determined by the nearby π-conjugated molecular layers. Since, irrespective of the specific microscopic transport mechanism, all experiments[18,19,21] and theoretical calculations[20,30,35] indicate that coupling of the carriers to a polarizable environment suppresses their mobility, we suggest that this intrinsic dielectric response of the semiconductor crystal plays a key role in determining which molecular semiconductors are more likely to show high mobility and band-like transport.

The influence (on the field-induced charge carriers) of the dielectric polarizability of the π-conjugated molecular layers in the organic semiconductor can be reduced by either increasing their distance (as discussed above for PDIF-CN$_2$) or by decreasing their polarizability. This second point is important because the polarizability tensor of individual rod-like aromatic molecules is anisotropic: it is larger along the longest molecular axis and smaller for the other directions.[35] Molecular orbital calculations for different molecules of interest in the field of organic electronics indicate that a factor of 2-3x difference in the polarizability for the different molecular directions is common.[36,37] It follows that also the macroscopic dielectric constant of organic crystals is anisotropic, and it depends on the crystal packing.[37] If the molecules are packed with their long axis perpendicular to the layer where charge carriers propagate, the suppression of mobility due to the dielectric constant of the nearby π-layer to the channel is the largest. However, when the packing is such that the long axis of the molecules lies in the transport layer, the effect of the semiconductor dielectric environment is minimized, resulting in improved transport characteristics. These two key factors (the presence of substituents, increasing the distance between the π-cores of the molecules and the orientation of the molecules themselves), which determine the coupling of the charge carriers to the dielectric environment, are schematically illustrated in **Fig. S2** of the supporting information.

The validity of these concepts is supported by recent measurements on *p*-channel single crystal OFETs (**Fig. 4**). After the discovery of band-like transport in rubrene, a similar, reproducible behaviour was observed for TMTSF[14] and $C_8$-BTBT[15] single crystal transistors as well as for thin films of TIPS-pentacene.[38] As a term of comparison, no mobility increase with decreasing temperature has been reported to date in FETs of pentacene and sexithiophene despite prolonged research (see also discussion in supporting information). These observations can be rationalized in terms of the ideas introduced above, by analyzing the effects of the semiconductor molecular polarizability. Specifically, as for PDIF-$CN_2$, in $C_8$-BTBT crystals the influence of the polarizability of the π-conjugated layers next to the transistor channel is reduced by the substituents connected to the π core. In rubrene, TMTSF, and TIPS-pentacene single crystals the packing is such that the long axis of the π-core lies in the transport plane (Fig. 4), reducing the coupling of the charge carriers to the polarizability of the organic semiconductor. On the other hand, if the long axis of the molecule is stacked perpendicular to the π-conjugated layer, the higher effective polarizability "seen" by the charge carriers will enhance the tendency to localization. This is the case of pentacene and sexithiophene, where band-like transport in FETs is not observed. Thus, to a first approximation, *the effect of the different molecular orientation is similar to that of a change in the dielectric constant of the gate insulator.*

These considerations allow us to conclude that differences – for different crystal structures – of 2-3x in the dielectric constant due to the orientation of the nearby π-conjugated layers, as well as differences in the interlayer spacing on the scale of 1 nm have sizeable effects on the temperature dependence of the mobility, and in determining whether band-like transport is observed. Indeed, simple electrostatics indicate that the effects that we are considering can increase the depth (in energy) of the localized states in the band tail of the organic semiconductor by several tens of meV – comparable or larger than $k_BT$ at room temperature – in a way similar to the one discussed above to describe the effect of the gate dielectric on the mobility of electrons in PDIF-CN$_2$. This conclusion is in line with theoretical studies that have analyzed different microscopic models and scenarios indicating how the polarizability of an organic material affects the carrier mobility, causing its reduction in all cases.[19,20,30,35] . While it is clear that more work remains to be done to understand in detail and quantify more precisely the effect of the polarizability due to the molecules themselves on the transport properties of high quality organic FETs, the ideas that we have presented here are important because they link in a very simple and transparent way the structure of the material (both of the constituent molecules and of their packing) to the behavior of the carrier mobility. These ideas can therefore be used as a very practical guide for future experiments (e.g., investigations of the carrier mobility in crystals of a same molecule with substituents of different length), to improve the transport properties of organic semiconductors even further for useful applications,[39,40] and for detailed microscopic theoretical modeling.[41,42]

*Acknowledgements*
The authors would like to thank D. Frisbie, who first brought to their attention the importance of the anisotropic polarizability of conjugated molecules and of their packing, in relation to the occurrence of band-like transport. We also thank J. Takeya for helpful discussions as well as for providing Cytop and A. Ferreira, I. G. Lezama, M. Nakano, B. Sacépé and H. Xie for technical assistance. The study was supported in part by the Swiss National Science Foundation, NEDO, and MaNEP.

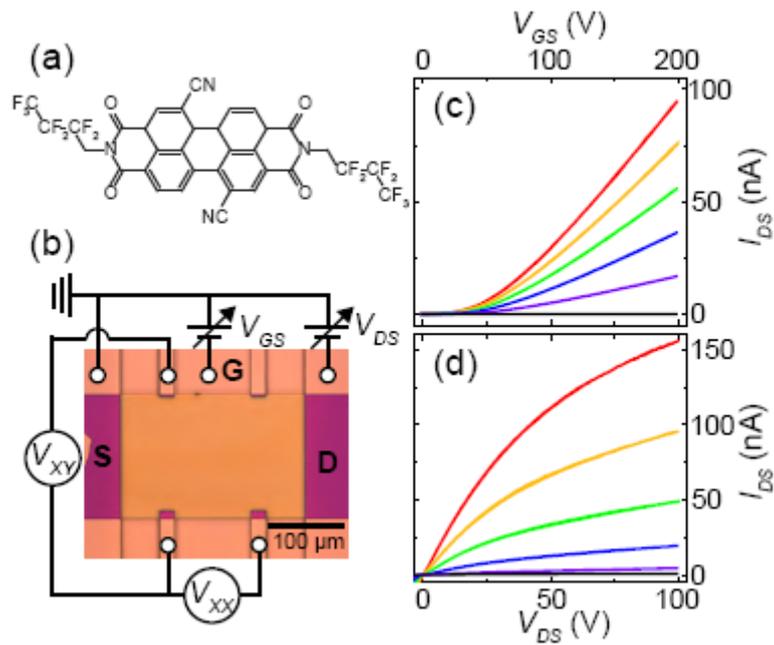

**Figure 1.** (a) Chemical structure of PDIF-CN$_2$. (b) Optical microscope image of a vacuum-gap device with the circuit schematics illustrating the configuration used for the four-terminal resistance measurements. **S** and **D** denote the source and drain contacts, and **G** the recessed gate. The voltage probe are also clearly visible (all contacts consist of an evaporated titanium/gold bilayer film) (c) Transfer characteristics of a vacuum-gap device with a 25 μm recessed gate (this distance is kept intentionally large to enable the insertion of an ionic liquid when desired[27]) which is why the value of the gate voltage applied is particularly large. In these measurements $V_{DS} = 0,2,4,6,8,10$ V, respectively. (d) Output characteristics of the same device ($V_{GS} = 0,20,40,60,80,100$ V).

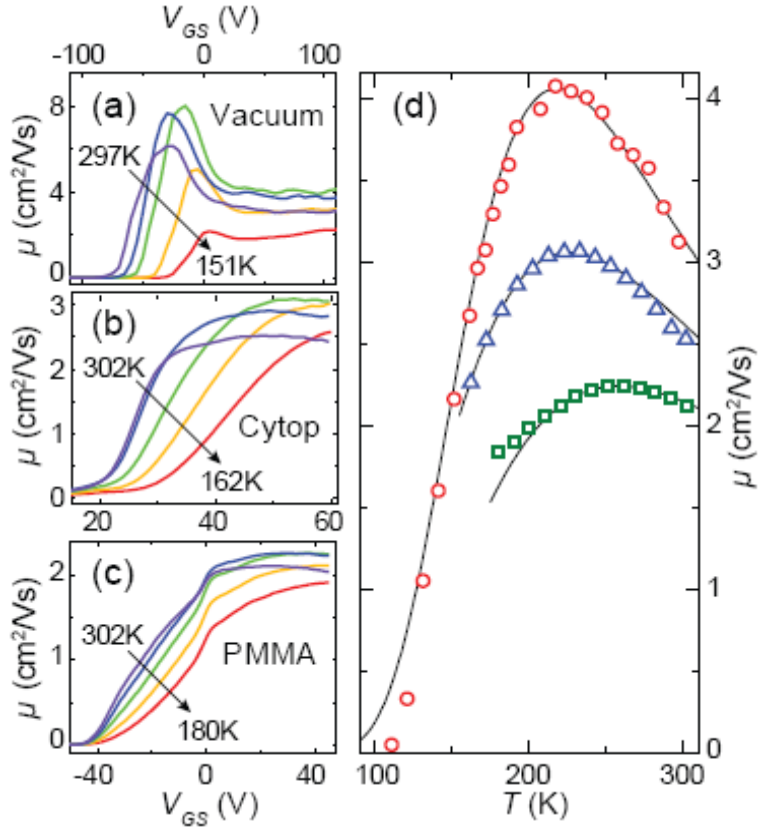

**Figure 2.** Panels (a)-(c) show the gate voltage dependence of the mobility of PDIF-CN$_2$ single-crystal OFETs with different gate dielectrics (vacuum, Cytop, and PMMA, respectively). The gate voltage range is different because of the different relative dielectric constant and thickness of the three insulators. Devices on Cytop and PMMA regularly break in a range of temperatures just below 200 K – probably because of differences in the thermal expansion coefficient of the substrates – and could not be measured at lower temperatures. (d) Temperature dependence of the mobility for devices on the three different dielectrics. The mobility values are determined at large gate voltage values, where $\mu(V_{GS})$ is approximately constant. The open circles, triangles, and squares correspond to data measured on vacuum-gap, Cytop, and PMMA devices, respectively. The continuous lines are fits to the model discussed in the main text.

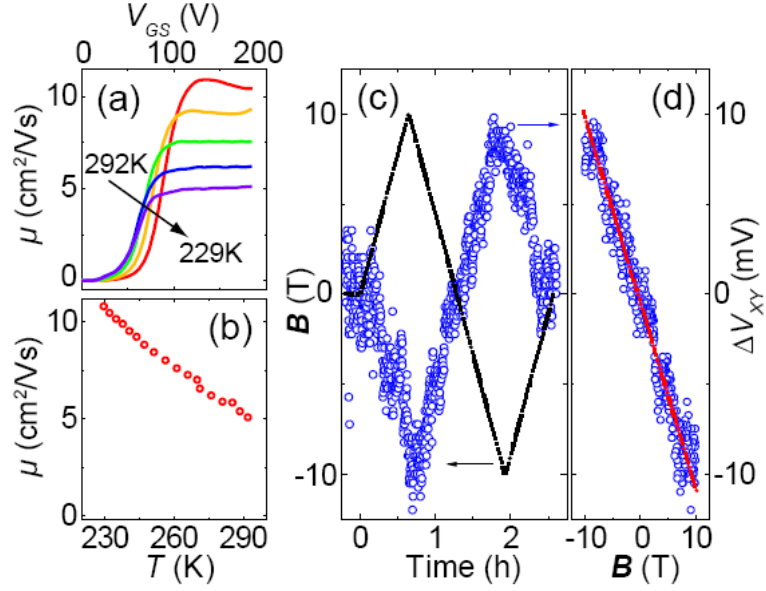

**Figure 3.** (a) Mobility extracted from the transfer characteristics as discussed in the text, as a function of gate voltage, for different temperature values (from 290 K to 230 K; the device broke while cooling down further). The temperature dependence of the mobility at high gate voltage is shown in panel (b): $\mu(T)$ increases from 5.1 cm$^2$/Vs at $T$ = 290 K to 10.8 cm$^2$/Vs at $T$ = 230 K. (c) The open circles show the time-evolution of the change in transversal voltage $\Delta V_{XY}$ (right axis) measured while applying constant $V_{DS}$ and $V_{GS}$ ($I_{DS} \approx 5.5 \cdot 10^{-8}$ A) and sweeping the magnetic field $\boldsymbol{B}$ at $T$ = 270 K (a magnetic field independent offset voltage due to misalignment of the voltage probes has been subtracted). The dotted line shows the applied magnetic field as a function of time. (d) Measured Hall voltage $\Delta V_{XY}$ plotted versus magnetic field. The red line is a linear fit, from the slope of which the carrier density is extracted.

## Molecular Structure

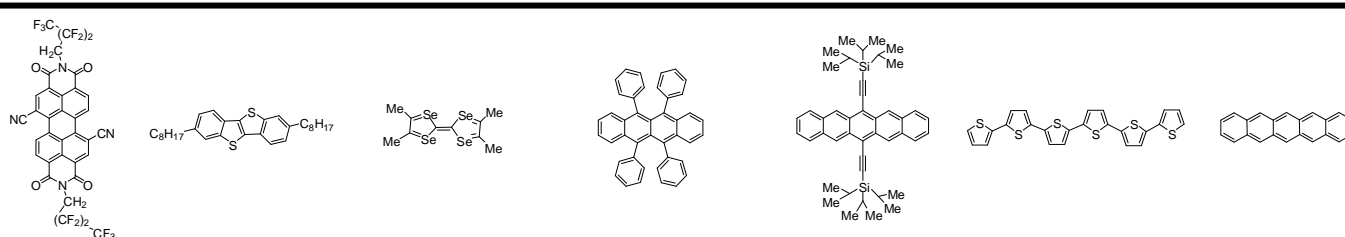

## Crystal Structure

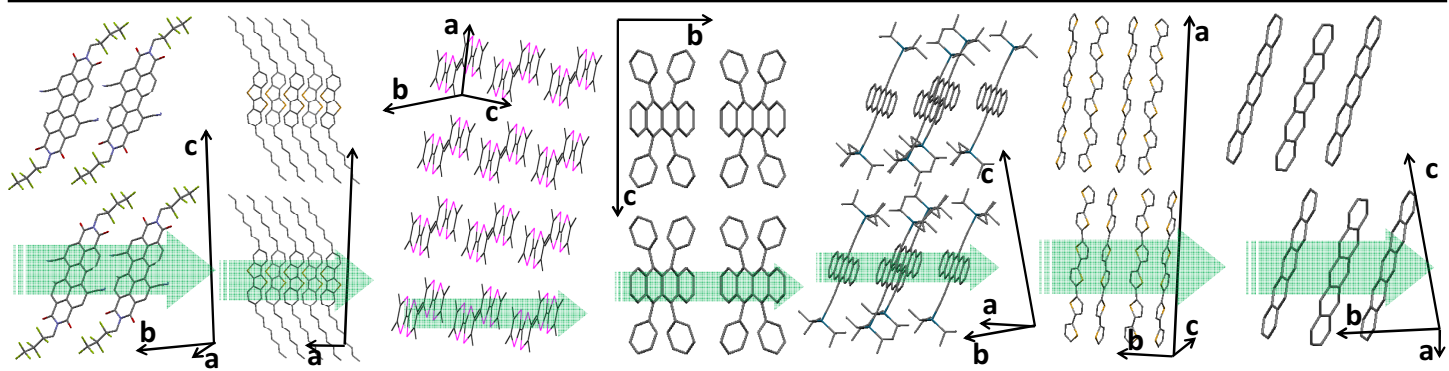

## Maximum Single-Crystal FET Mobility (cm$^2$/Vs)

| | | | | | | | |
|---|---|---|---|---|---|---|---|
| **HT** | 6.0 | 9.1 | 4.0 | 20 | 1.8 | 0.13 | 2-3 |
| **LT** | 10.8 | 11.0 | 6.0 | 35 | 6.0 | - | - |

**Figure 4.** Analysis of the molecular structure and packing for different organic molecules (from left to right: PDIF-CN$_2$, C$_8$-BTBT, TMTSF, rubrene, and TIPS-pentacene, sexithiophene, pentacene). To date, PDIF-CN$_2$, C$_8$-BTBT, TMTSF, rubrene, and TIPS-pentacene are the only organic semiconductors exhibiting band-like transport in an OFET configuration. They satisfy the criteria discussed in the main text to minimize the coupling between charge carrier and molecular polarizability. For sexithiophene and pentacene – that do not satisfy any of these criteria – band-like transport in a FET configuration has never been reported, despite the considerable research efforts that have focused on these molecules. The values of mobility cited are either at room temperature (HT) or low temperature (LT – for the molecules in which band-like transport has been observed). Additional details about mobility values of different materials are discussed in the supporting information.

*Supporting Information:*

**Experimental**

PDIF-CN$_2$ single crystals were grown by physical vapor transport in a flow of argon gas. Vacuum-gap devices were fabricated by manually laminating crystals onto elastomeric poly(dimethylsiloxane) (PDMS) substrates[1] with Ti/Au source and drain contacts and a 25 μm recessed gate. Devices with Cytop or PMMA as gate dielectrics were fabricated by spin-casting onto Si/SiO$_2$ substrates followed by baking on a hotplate.[2] Ti/Au contacts were deposited through shadow masks in this case and the capacitance values were measured in a parallel-plate configuration using an Agilent Technology 4284A Precision LCR meter. The length and width of the laminated single crystals typically range from 100 to 500 μm with a thickness of a few microns. All devices were first characterized by measuring the transfer and output characteristics in vacuum ($p \approx 5 \cdot 10^{-7}$ mbar) at room temperature (RT) using an Agilent Technology E5270B parameter analyzer. An optical image together with the transfer and output characteristics of a vacuum-gap device (forward and reverse sweeps) are shown in Fig. 1; the corresponding images and characteristics for devices with Cytop and PMMA as gate dielectric are shown in Fig. S1, demonstrating the high quality of the devices. Variable temperature measurements were carried out in a helium flow cryostat; the sample temperature was measured on the cold finger next to the sample holder. Hall measurements were performed at constant drain-source and gate-source voltages $V_{DS}$ and $V_{GS}$ in perpendicular magnetic field in a $^4$He cryostat with a variable temperature insert; the drain-source current was $I_{DS} \approx 5.5 \cdot 10^{-8}$ A.

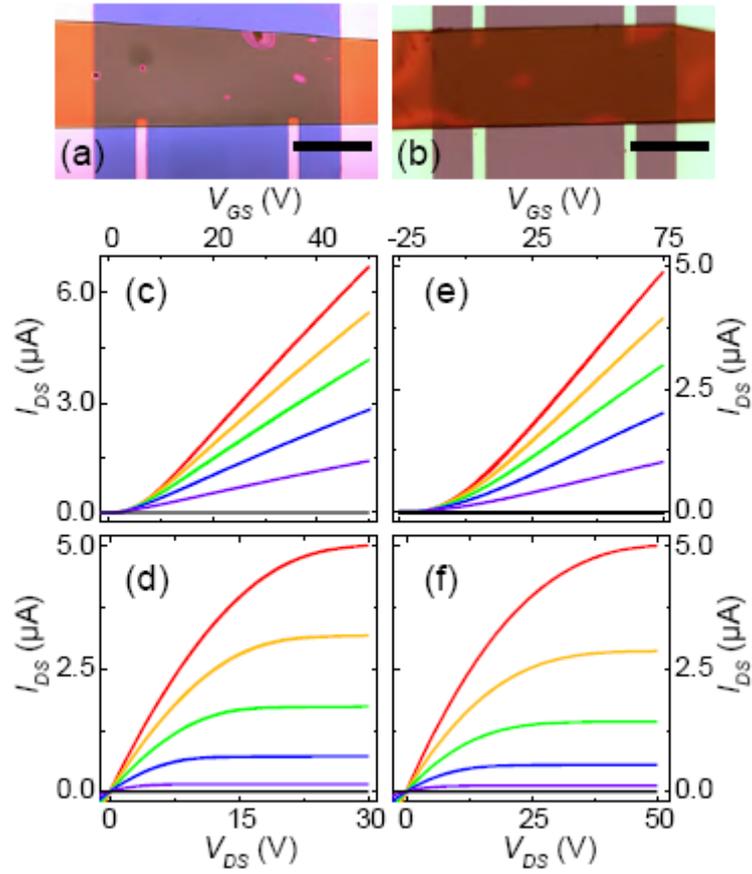

**Figure S1.** Single-crystal PDIF-CN$_2$ OFETs on polymeric gate insulators. Panels (a) and (b) show optical microscope images of single-crystal PDIF-CN$_2$ devices on doped silicon wafers coated with Cytop (a) and PMMA (b) layers, acting as gate dielectrics. The Ti/Au contacts were evaporated through a shadow mask; the scale bar is 100 µm. (c) Transfer and (d) output characteristics of a device with Cytop gate dielectric ($V_{DS}$ = 0,2,4,6,8,10 V in (c) and $V_{GS}$ = 0,5,10,15,20,25 V in (d)). Panels (e) and (f) show the same measurements on a device with PMMA gate dielectric (the data is for the same sequence of $V_{DS}$ and $V_{GS}$ values, as shown in panels (c) and (d)).

**Modeling of $\mu(T)$**

To estimate the role of the dielectric dipolar disorder in the band-like transport regime we use a simple phenomenological model that we have developed to analyze the behavior of TMTSF single-crystal FETs.[3] In highly-ordered molecular crystals, the states relevant for FET conduction consist of a band in which electrons move with an intrinsic mobility (states at energy $E > 0$), and of a band tail (states at energy $E < 0$) caused by the remnant disorder in the crystal, in which electrons are localized ($E=0$ separates localized and delocalized states). The density of states in the band of PDIF-CN$_2$ of is taken to be $N_0 = 10^{15}$ cm$^{-2}$eV$^{-1}$ (corresponding to the density of molecules at the surface divided by the bandwidth, taken to be ~0.5 eV) for energy $E > 0$; disorder-induced trap states are represented by a Gaussian-like distribution of tail states $N_t(E) = N_t \cdot \exp(-E^2/2E_t^2)$ for $E < 0$, where $N_t$ and $E_t$ are the height and width of the distribution (other distributions lead to the same result[3,4]).

Under these assumptions, it is straightforward to calculate the density $n_b$ of electrons in the conduction band for any given gate voltage and temperature and, consequently, the FET mobility. The drain-source current as a function of temperature $T$ and total density of induced charge $n = C_i \cdot (V_{GS} - V_{Th})/e$ is given by $I_{DS}(n,T) = n_b(n,T)e\mu_0(T)V_{DS}W/L$. Therefore, the mobility is determined by $\mu(T) = \mu_0(T) \cdot \partial n_b / \partial n \propto \partial I_{DS}(T)/\partial V_{GS}$, where the intrinsic mobility is assumed to depend on temperature as $\mu_0(T) = \alpha \cdot T^{-2}$ [5,6,7] with a material-specific parameter $\alpha$.

In practice, we calculate the position of the Fermi energy $E_F(n,T)$ numerically by solving the equation $n = n_b + n_t = \int_0^\infty N_0 \cdot \left(e^{E-E_F/kT} + 1\right)^{-1} dE + \int_{-\infty}^0 N_t(E) \cdot \left(e^{E-E_F/kT} + 1\right)^{-1} dE$ for a given density of states. Having determined $E_F(n,T)$, we calculate $\mu(T)$ in terms of the density of charge carriers occupying band states $n_b = \int_0^\infty N_0 \cdot \left(e^{E-E_F/kT} + 1\right)^{-1}$ as described above. Setting a common value for $\alpha$ in the three cases (i.e., for devices with vacuum gap, PMMA and Cytop as gate dielectric; $\alpha$ is a material parameter that in principle depends only on the properties of PDIF-CN$_2$) leaves two fitting parameters $N_t$ and $E_t$ which determine the height and width of the band tail. It is found that the smaller the width of the Gaussian ($E_t$), the closer the Fermi level is to the edge of the band of the delocalized states, and the larger is the temperature range over which the mobility increases upon lowering the temperature (band-like transport occurs when $E_t$ is comparable to or smaller than $k_BT$). The parameters $N_t$ and $E_t$ extracted from the fits to the experimental data are 38.0 meV and $2.4 \cdot 10^{13}$ cm$^{-2}$eV$^{-1}$ for vacuum as gate dielectric, 42.7 meV and $3.8 \cdot 10^{13}$ cm$^{-2}$eV$^{-1}$ for Cytop, and 55.9 meV and $2.9 \cdot 10^{13}$ cm$^{-2}$eV$^{-1}$ for PMMA, respectively.

**Influence of the molecular structure on charge transport**

Fig. S2 schematically illustrates the different contributions to the polarizability that affect the charge carriers accumulated at the surface of the organic crystal, in a FET configuration, as discussed in the main text. For air-gap devices, vacuum acts as gate dielectric, and the dielectric environment experienced by the charge carriers is entirely determined by the crystalline molecular layers in the organic organic crystal (i.e., the planes of molecules in the organic crystal next to the channel). The strength of the coupling between charge carriers and this dielectric environment is determined by the molecular structure and crystal packing. Key factors are the presence of substituents, increasing the distance between the π-cores of the molecules (the longer the substituent, the weaker is the coupling) and the orientation of the molecules themselves, whose anisotropic polarizability (indicated by the yellow arrows in the figure) determines the dielectric constant of the nearby planes that is experienced by the charge carriers (this dielectric constant plays a role analogous to the dielectric constant of the gate dielectric in non-suspended devices).

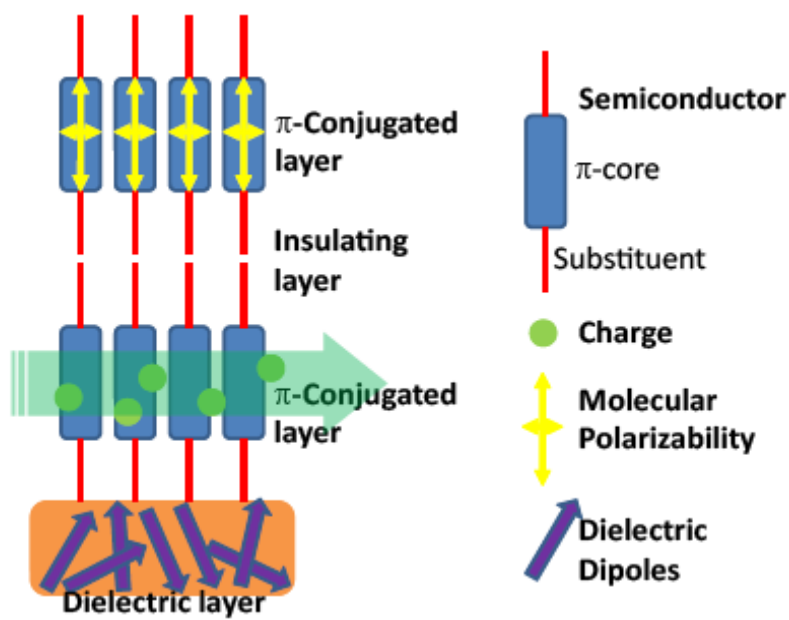

**Figure S2.** Influence of the molecular structure on charge transport in an OFET accumulation layer. Schematic representation of the key ingredients determining the dielectric environment experienced by the charge carriers propagating in the channel of an organic field-effect transistor. The role of the gate dielectric (illustrated in the figure by the dielectric dipoles) has been well-established experimentally for *p*-channel devices in recent work; similar effects (albeit of smaller magnitude) for *n*-channel OFETs are reported in this work. In the absence of the gate insulator (i.e., for air-gap devices), the dielectric environment is entirely due to the crystalline molecular layers in the organic semiconductors, next to the channel. As discussed in the main text, the strength of the coupling to this dielectric environment is determined by the molecular structure and crystal packing. Key factors are the presence of substituents increasing the distance between the π-cores of the molecules, and the orientation of the molecules themselves, that play a role through their anisotropic polarizability (indicated by the yellow arrows in the figure).

**Mobility values for different molecules**

Among all the materials that we have considered in the main text as part of the discussion of structure-property relations, several have been investigated in our group using vacuum-gap transistors. These include PDIF-CN$_2$, TMTSF, Rubrene, and pentacene. For all these materials we have successfully realized high-quality devices that reproducibly show no hysteresis, are stable over time (at least as long as the devices are kept in vacuum), and exhibit gate voltage independent mobility. In these investigations we have accumulated sufficient statistics, which enables us to make clear statement about the properties of OFETs based on these materials. We want to make clear that, despite the use of single-crystals, these values may not represent the ultimate mobility values for each material. Residual disorder is likely present and at this time there are no established characterization techniques that allow quantifying the amount and sources of disorder (nor what is the precise effect on the carrier mobility) with sufficient precision. Nevertheless, the results obtained on single crystal FETs investigated in our group have been reproduced by other groups, sometimes under identical settings (e.g., rubrene and pentacene, by Podzorov's group), and in other cases under slightly different conditions (this is the case for TMTSF and PDIF-CN$_2$), which still allow us to conclude that the mobility values reported are representative of the quality of existing top devices. For OFETs based on other materials (C$_8$-BTBT, TIPS-pentacene, and thiophene) we have referred to the literature, taking care to consider results that have been reproduced by more than one group.

These considerations are important, especially because going through the literature it is easy to find mobility values quoted – in certain cases, very large – that are not backed up by unambiguously clear transport data obtained from a sufficiently large number of devices (i.e., insufficient statistics), or which have been obtained by operating the devices in particular conditions. This is the case, for instance, for mobility values that are strongly gate voltage dependent: since in the (low) carrier density regime in which conventional OFETs are operated there is no intrinsic reason to have a density dependence of the mobility, the observation of a strongly gate voltage dependent mobility indicates that some specific phenomenon is influencing the device characteristics[8] (accordingly, such values are not representative of the general behavior). Certainly, the case of pentacene is indicative, as many groups have reported mobility values varying over a broad range both in thin films OFETs and in single crystals,[9] and the reasons for this spread – probably originating from differences in the processing – have never been pinpointed in detail (with the largest values reported not having been reproduced by independent groups). Our group[10] and the group of Podzorov[11] have independently and reproducibly obtained mobility values of 2-3 cm$^2$/Vs at room temperature in vacuum-gap devices. These values, obtained by two different groups using similar fabrication protocols is – on the contrary – indicative of good reproducibility in different laboratories.

We conclude that, while reports of high mobility OFETs are certainly interesting, they have to be taken with care before considering them as representative of top-quality devices. Table S1 summarizes the main references that we have considered in assessing what is the top quality (and the highest representative mobility values) of transistors realized using the materials discussed in this paper. *The maximum mobility values stated for the different molecules in Fig. 4 were selected to the best of our knowledge based on the requirement that a similar value was reported by at least two independent groups*[9-29].

**Table S1.** Comparison of field effect mobility values for organic molecular crystal discussed in the article.

| Molecule | Processing Method | Gate dielectric | $\mu^{FET}$ (RT)[a] (cm$^2$/Vs) | $\mu^{FET}$ (T*)[b] (cm$^2$/Vs) | T* (K)[c] | Ref. |
|---|---|---|---|---|---|---|
| PDIF-CN$_2$ | PVT | vacuum | 5.1 | 10.8 | 230 K | current work |
|  | PVT | PMMA | 6.0 |  |  | 12 |
|  | PVT | Cytop | 3 |  |  | 13 |
| C$_8$-BTBT | Solution | PMMA | 9.1 | 11.0 | 100 K | 14 |
|  | Solution | SiO$_2$ | 5.0 |  |  | 15 |
|  | Solution | SiO$_2$ | 3.5 |  |  | 16 |
| TMTSF | PVT | vacuum | 4.0 | 6.0 | 160 K | 17 |
|  | PVT | Parylene | 4.0 |  |  | 18 |
| Rubrene | PVT | vacuum | 20 | 30 | 180 K | 19 |
|  | PVT | vacuum | 20 | 35 | 180 K | 20 |
| TIPS-Pen | solution | PI | 1.2 | 6.0 | 4.3 K | 21 |
|  | solution | SiO$_2$/HMDS | 1.8 |  |  | 22 |
| 6-T | PVT | PMMA | 0.075 |  |  | 23 |
|  | thin film | Parylene | 0.13 |  |  | 24 |
| 8-T | thin film | SiO$_2$ | 0.33 |  |  | 25 |
| DD-6T | thin film | PVP | 0.5 |  |  | 26 |
| Pentacene | PVT | SiO$_2$ | 2.3 |  |  | 27 |
|  | PVT | Parylene | 2.2 |  |  | 28 |
|  | PVT | vacuum | 2.0 |  |  | 10 |

| | | | |
|---|---|---|---|
| PVT | vacuum | 3 | 11 |
| PVT | Parylene | 0.3 | 29 |

[a]Field-effect mobility at room temperature. [b]Maximum field-effect mobility measured at low temperature. [c]Temperature, down to which the field-effect mobility increased. PVT = physical vapour transport. SAM: Self-assembled monolayer. PI: Polyimide. HMDS: Hexamethyldisiloxane. PVP: Poly-4-vinylphenol.

**References (Supporting Information)**